\documentclass[final] {aipproc}
\usepackage{sidecap,graphicx,boxedminipage,epsfig,wrapfig,floatflt,shadow}

\layoutstyle{8x11double}

\begin{document}

\title{Helping Students Learn Quantum Mechanics for Quantum Computing}


\author{Chandralekha Singh}{
  address={Department of Physics and Astronomy, University of Pittsburgh, Pittsburgh, PA, 15260, USA}
}

\begin{abstract}
Quantum information science and technology is a rapidly growing interdisciplinary field drawing researchers from
science and engineering fields. Traditional instruction in quantum mechanics
is insufficient to prepare students for research in quantum computing because there is a lack of emphasis in
the current curriculum on quantum formalism and dynamics.
We are investigating the difficulties students have with quantum mechanics and 
are developing and evaluating quantum interactive learning tutorials (QuILTs) to reduce the difficulties.  
Our investigation includes interviews with individual students and the 
development and administration of free-response and multiple-choice tests. 
We discuss the implications of our research and development project on helping students learn
quantum mechanics relevant for quantum computing.
\end{abstract}

\maketitle

\section{Background}
\vspace*{-.09in}
Quantum computing is a rapidly growing interdisciplinary area of research involving researchers from 
physics, chemistry, electrical engineering, computer science and engineering, and material science disciplines~\cite{book}. 
Feynman was the first to realize that quantum systems performed computations that might be exploitable for large
scale computing~\cite{feynman}. In 1994, Shor~\cite{shor} developed a powerful and efficient algorithm to factor prime 
numbers on a quantum computer which is exponentially faster than the classical algorithms.
The importance of Shor's algorithm to national security instantly started 
a race to develop a ``real" scalable quantum computer because the difficulty in factoring
large prime numbers is at the heart of the protocols used for encoding/decoding secret information. 
The encoding and decoding protocols that rely on the inability of the hackers to factor large prime numbers 
are also responsible for most secure communication, {\it e.g.}, credit card 
transactions over the internet. Feynman's vision from more than 25 years ago that 
quantum mechanics should be exploited to perform
fast computing has come alive with government agencies investing large resources into quantum computing technologies.
Unfortunately, the quantum mechanics curriculum in various departments is not suited to prepare students
for research in quantum computing.
For those interested in quantum algorithms, learning the quantum mechanics formalism is easier than 
for those involved in the experimental realization of quantum computers. 
While the first group must have a good grasp of the quantum formalism for a two level system and product states,
the latter group must also consider practical issues involved in experimentation. 

Quantum infomation is stored in quantum bits (qubits). Unlike a classical bit which can only
take two values (0 and 1), a qubit can be in a quantum superposition of $\vert 0 \rangle$ and $\vert 1 \rangle$:
$\vert \Psi \rangle =\alpha_0 \vert 0 \rangle
+\alpha_1 \vert 1 \rangle$ where the only constraints on the complex coefficient is that 
$ \vert \alpha_0 \vert^2+\vert \alpha_1 \vert^2=1$. 
For n-qubit system, $2^n$ complex numbers are required. For example, for two qubits, 
$\vert \Psi \rangle =\alpha_0 \vert 00 \rangle
+\alpha_1 \vert 01 \rangle +\alpha_2 \vert 10 \rangle +\alpha_3 \vert 11 \rangle $.
A state with $n=100$ qubits is specified by $2^{100} \sim 10^{30}$ coefficients! A quantum
program is specified by ${(2^{100})}^2= 10^{60}$ coefficients but the final answer is a string of $n=100$ classical bits.

DiVincenzo~\cite{divincenzo} has put forward these 
five criteria for solid state 
implementation of a quantum computer:
\begin{itemize}
\item Scalable physical system with well-defined qubits
\item Be initializable to a simple state such as $|000...>$
\item Have much longer decoherence times than computation time
\item Have a universal set of quantum gates
\item Permit high quantum efficiency, qubit-specific measurements
\end{itemize}

As can be seen from these criteria, the practical issues in building a ``scalable" quantum computer 
include challenges in making an actual qubit considering most quantum systems will have more than two levels,
issues related to state preparation ({\it e.g.}, for initializing the register at the start of a computation),
making real quantum gates (which involves practical issues related to the time evolution of a quantum state), 
minimizing decoherence in the system, and performing efficient measurements to read the output of the computation.
It can be shown that two qubit gates are universal for quantum computation~\cite{divincenzo}.
\vspace*{-.20in}
\section{Research Objectives and Methodology}
\vspace*{-.10in}
We have been carrying out research on the types of difficulties students have with the formalism of quantum mechanics
and developing Quantum Interactive Learning Tutorials (QuILTs) to reduce the difficulties~\cite{my}. Issues related to state preparation
({\it e.g.}, for initializing a quantum computer), time development (for making quantum gates for performing the actual computation), 
measurement (for reading the output of a computation), and basics of two level systems ({\it e.g.}, spin one-half) and product space 
are some of the topics we are targeting. In the following section, we briefly describe some
of the findings.

The research methodology involves administering written surveys to advanced undergraduate students and beginning graduate students.
In these written surveys, students were asked to explain their reasoning. In addition, we also conducted individual interviews
with students using a think-aloud protocol. In these interviews, we initially allowed students to answer the questions posed 
to the best of their ability without interruption and then probed them further about issues they did not otherwise make clear.
Many of the probing questions were developed ahead of time, but some were generated on-the-spot in light of student responses.
\vspace*{-.2in}
\section{Difficulty with quantum measurement}
\vspace*{-.1in}
Students were posed the following question:
``Consider the following conversation between Andy, Caroline, and John
about the measurement of an observable $A$:\\
$\bullet$ Andy: When an operator $\hat A$ corresponding to a physical observable $A$
acts on a wave function $\Psi $, it corresponds
to a measurement of that observable. Therefore, $\hat A \Psi =\lambda_a  \Psi $.\\
$\bullet$ Caroline: I disagree. The measurement collapses the wave function
so $\hat A  \Psi =\lambda_a \Psi_a $ where $\Psi_a $ is an eigenfunction of $\hat A$. \\
$\bullet$ John: I disagree with both of you. You cannot represent the instantaneous collapse of a wave function upon
the measurement of $A$ by either equation. Rather, you can write the wave function right before the measurement
as a linear superposition of the eigenfunctions of $\hat A$, {\it i.e.}, $\Psi =\sum_a \beta_a \Psi_a $.
Then, the absolute square of the coefficients $\vert \beta_a \vert^2$ give the probability of collapsing into
$\Psi_a $ and measuring $\lambda_a$.\\
$\bullet$ Andy: Then, what is $\hat A \Psi =$?\\
$\bullet$ John: $\hat A$ acting on $\Psi $ is not a statement about the measurement of $A$. Rather,
$\hat A \Psi = \hat A \sum_a \beta_a \Psi_a = \sum_a \lambda_a \beta_a \Psi_a $. \\
With whom do you agree? Explain why the other two are not correct."\\
\vspace{0.06in}
John's statement is correct. Surprisingly, many interviewed students incorrectly stated that both Caroline and John are actually saying the same thing
and they are both correct despite the fact that John explicitly says that he disagrees with the other two.
Then, students were explicitly asked to explain how a linear combination of the eigenfunctions of $\hat A$ that John proposes
can be the same as only one term in the sum proposed by Caroline in $\hat A \Psi =\lambda_a \Psi_a $.
Most of these students explained their reasoning by claiming that the Hamiltonian operator acting on the wave function
corresponds to the measurement of $A$ as Caroline proposes. They incorrectly added that John's equation 
$\hat A \Psi = \hat A \sum_a \beta_a \Psi_a = \sum_a \lambda_a \beta_a \Psi_a $
is true only before the measurement of $A$ has actually taken place
and Caroline's statement $\hat A  \Psi =\lambda_a \Psi_a $ is true right after the measurement of $A$ has taken place
and lead to the collapse of the wave function. Many students explicitly stated that right at the instant the measurement takes
place both Caroline and John are correct because the wave function undergoes an instantaneous collapse and the 
right-hand-side (RHS) of the equation changes. 

When the interviewed students were explicitly asked how the RHS of an equation can change when the left-hand-side (LHS)
remains the same, many students appeared not to be concerned about such an anomalous situation in linear algebra where depending
upon the context, the same LHS yields different RHS. Students were often very focused on the context.
They were convinced that the collapse of a wave function upon the measurement of an observable
in quantum mechanics must be represented by an equation and Caroline's equation must correspond to the equation after the collapse
of the wave function has occured.
They often reiterated that such changes occur only
to the RHS (and the LHS is the same for both John and Caroline) because RHS corresponds to the ``output" 
and the LHS corresponds to the ``input". According to their reasoning, it is only the output that is affected by the measurement
process (and not the input) so the LHS for John and Caroline are the same.
When students were asked to explicitly choose the observable to be energy so that the operator is the Hamiltonian operator,
their qualitative responses were unchanged even in that concrete case.

The above example shows how difficult the quantum measurement postulate based upon the Copenhagen interpretation is
and how students have built a locally coherent knowledge structure (inconsistent with the quantum postulate) to represent the 
measurement process with equations. 
It is also interesting to note that since students were often convinced about the physical process of the
wave function collapse as represented by the equations that John and Caroline wrote (related to $\hat A$ acting on $ \Psi $),
they blurred out the linear algebra involved and did not question the anomaly regarding the same LHS yielding different RHS.
We plan to administer a modified version of the question to students who have taken linear algebra but not 
quantum mechanics. Students can be asked to explain why they agree or disagree with a person who says that
a physical process can change the equation written by John to that written by Caroline ({\it i.e.}, the LHS of the equation
remains the same but the RHS changes). Our hypothesis is that in the absence
of the knowledge of the ``collapse" hypothesis and an attempt to represent the collapse by an equation, students who know linear algebra
will agree with John and argue that Caroline's equation does not make sense. 
           
Written tests and interviews suggest that students have difficulty figuring out what the wave function
will be at time $t$ after the measurement of a physical observable.
Many students believe that after the measurement of \textit{any} 
observable, the system gets \textquotedblleft stuck\textquotedblright\ in
the eigenstate of the corresponding operator forever unless an external perturbation is applied. 
For example, many students believe that the wave function continues to be a position eigenfunction
after the measurement of position of a quantum mechanical particle because an eigenfunction cannot change with time.
Of course, the statement is true only for observables whose operators commute with the
Hamiltonian, but students seem to have overgeneralized this property to include all observables. 

Incidentally, when asked to plot an example, many students do not know what a position eigenfunction may look like. 
During the interviews, students were asked to plot a position eigenfunction 
on a $\Psi(x)$ vs. $x$ graph but such explicit instruction also did not help.
Written tests and interviews suggest that many students do not understand the meaning of ``an eigenfunction of an operator corresponding
to a physical observable" and believe that the eigenfunctions of all observables are the
same as the energy eigenfunctions. In interviews, many students explicitly stated that eigenfunctions do not evolve in time. 
When they were asked if a delta function in position is an
eigenfunction of any physical observable, some students said that it cannot be an eigenfunction because
it evolves in time and does not remain a delta function forever. The following response from
a student is a typical response: ``Energy eigenfunctions must be related to momentum and position
eigenfunctions...they are all eigenfunctions after all...shouldn't they at least be proportional to each other?"
One reason for such misconception
is that energy eigenfunctions which are emphasized in the 
courses are often simply called ``eigenfunctions".

When students were asked to write an eigenvalue equation for the position operator, many students had great
difficulty.
In the interivews, if students had difficulty writing an eigenvalue equation for the position operator, they
were then asked to write an eigenvalue equation for any operator. Roughly half
of the students wrote the Time-Independent Schroedinger Equation (TISE) which is an eigenvalue equation for the Hamiltonian operator but the
other half could not come up with anything reasonable. 
Even when prodded to recall a
general eigenvalue equation for a generic operator from a math course, they often
only recalled that there was a $\lambda$ involved (perhaps because they were asked to write an ``eigenvalue" equation).
Some claimed that TISE is not an eigenvalue equation when explicitly asked about it.

While many students believe that a quantum system gets stuck in an eigenfunction after a measurement,
a large number of students believe the opposite, {\it i.e.}, if one waits long enough, the time-evolution will
guarantee that the wavefunction after the measurement will go back to the 
``original" wavefunction (right before the measurement took place). As one student put 
it: ``...well it may not happen immediately
but if you wait for a sufficiently long time, it has to go back to the wavefunction before measurement."
Incidentally, this notion of going back to the original state was somewhat more prevelant if the wavefunction
before the measurement was the ground state wavefunction but it was also quite common when the wavefunction
right before the measurement was a linear superposition of the ground state and first excited state wavefunctions.
In the case of the ground state wavefunction, students often provided the justification that since the ground
state is the equilibrium state, the system must go back to it eventually. In the interviews, the interviewer told students
to consider the system to be completely isolated from the environment but very few students felt the need to re-evaluate
their claims that the system will go back to the ``original" state if one waits long enough. 
One student described the original state as the ``home" state and said
that after the collapse, the wavefunction has to find its home state eventually. When asked to show how the wavefunction
will evolve from the collapsed state to the ``home" state, the student added: ``I do not remember the 
calculation but the wavefunction's goal is to somehow get to the home state."
Some students with this belief felt that the collapse of the wavefunction upon measurement is
a mathematical construct and they only half-heartedly believed that the collapse can actually change
the wavefunction permanently.

On further prodding, 
responses of the interviewed students about their views on what should happen to the wavefunction a
time $t$ after the measurement is also intriguing.
When the interviewed students who believed that the wavefunction must go back
to the ``original one" 
were told that they should reconsider their response because their initial response is not correct, many
quickly switched to the notion that the wavefunction must then get stuck in the collapsed state.
When they were told that neither of these possibilities is correct, many students
responded in a manner similar to that of the following student: "aren't you contradicting yourself?" 
A similar situation occured when students who initially said that the system will get stuck in the
collapsed state were told that they should reconsider their response. After this
hint, many students promptly said that the system must go back to the original state. Stating
that neither of these responses is correct and asking students to reconsider their responses 
again made many students feel that the interviewers were contradicting themselves.

Thus, many of the advanced undergraduates and graduate students interviewed believed that there are only
two possibilities for the wavefunction: being stuck in the collapsed state or going back to the original
state before the measurement took place. They just could not contemplate the actual situation in which the
wavefunction evolves according to the Time-Dependent Schroedinger Equation (TDSE) and may neither be ``stuck"
(unless the state in which the wavefunction collapsed is an eigenfunction of an operator that commutes with
the Hamiltonian) nor ever go back to the ``original" state. 
When students were told that after the position measurement the
wavefunction is a delta function in position about a particular position and they were
explicitly asked to calculate the wavefunction after a time $t$, none of the students were able to perform the calculation.
The calculation involves expanding the delta function in terms of a linear superposition of eigenfunctions
and then computing the wavefunction at time $t$ by introducing appropriate phase factor $e^{-iE_nt/\hbar}$ to each term.
After the interviews, some of the students were very surprised to learn that in a majority of cases, the
wavefunction will never go back to the ``original" wavefunction if allowed to evolve according to TDSE.
Studying the real part of the wave function $\Psi(t)$ at different times $t$ via a suitable
simulation can convince students that the wavefunction
is neither ``stuck" in position eigenfunction nor must it go back to the original state. Such simulations can help students
learn that the time evolution operator can change the position eigenfunction significantly because the phase factors
$e^{-iE_nt/\hbar}$ of each term when the delta function is expanded in terms of energy eigenfunctions will evolve differently.

\vspace*{-.20in}
\section{Difficulty with Product Space}
\vspace*{-.10in}

When students were given two spin one-half particles and asked to choose a basis and write down a Hamiltonian 
$\hat H$ proportional to $\vec S_1 \cdot \vec S_2$ describing this system in a matrix form, a majority of the advanced undergraduates 
who had learned about product space had great difficulty. More than $85\%$ of students tried to construct a $2\times 2$ matrix because they
did not realize that they should consider a product space which is four dimensional. In the interviews, when students were specifically 
told that 
the product space of two spin one-half particles cannot be two dimensional, many of them remembered that the vector space 
should be four dimensional. Despite this realization, none of these students could actually choose a basis set and construct the Hamiltonian
correctly. Some claimed that $\hat H$ must always be diagonal regardless of the basis because it is the ``unperturbed" Hamiltonian (survey did
not mention it was the unperturbed Hamiltonian).
Some students had a ``cancellation" model in mind and they felt that if $\vec S_1$ and $\vec S_2$ are operators for two
spins, they have to conspire together to make the total spin of the system zero. 
For example, one student said: ``If the contribution
of $S_1$ is positive then the contribution of $S_2$ will be negative because their contributions must cancel."
They were often unable to articulate their reasoning.

Written tests and interviews suggest that many students have difficulty understanding that 
if two operators are in different Hilbert spaces, they will
always commute and can be treated independently of each other,
{\it e.g.}, spin and position of an electron or spins of two different electrons or positions of two different
electrons. Because of this difficulty, students often have difficulty figuring out how the operators
in different Hilbert spaces act on states in a product space.
For example, for the above Hamiltonian, several students felt that $\hat S_{1x}$ and $\hat S_{2y}$ will not commute with each other.

Our research also shows that determining the dimensionality of a product space is challenging for students.
For example, for a two spin system, students have a tendency to add (instead of multiply) the dimensionality of the individual
Hilbert spaces of each spin to obtain the dimensionality of the product space. This process will not introduce an error for
two spin one-half particles because $2+2=2\times 2$ but it will introduce error for other cases. For example, we have
found that for two spin-one particles, many students believe that the product space is 6-dimensional as opposed to 9-dimensional.
           
\vspace*{-.22in}
\section{Conclusion}
\vspace*{-.10in}

Research on student understanding of aspects of quantum mechanics relevant for quantum computing is necessary. We have
been investigating the difficulties students have in learning quantum mechanics and developing Quantum Interactive
Learning Tutorials (QuILTs) to reduce the difficulties. The tutorials can provide scaffolding support to students
in science and engineering pursuing quantum computing research.

\vspace*{-.24in}
\begin{theacknowledgments}
We are grateful to the NSF for award PHY-0244708.
\end{theacknowledgments}
\vspace*{-.20in}

\bibliographystyle{aipproc}

\end{document}